\documentclass[usenatbib]{mn2e}
\usepackage[dvips]{graphicx}
\usepackage{epsfig}
\usepackage{lscape}
\usepackage{amssymb,amsmath}

\newcommand{\kms}{{\rm \,km\,s^{-1}}}
\newcommand{\Mpc}{\,{\rm Mpc}}

\newcommand{\msub}{m_{\rm sub}}

\newcommand{\Rcusp}{R_{\rm cusp}}

\title{Substructure Lensing: Effects of Galaxies, Globular Clusters \&
Satellite Streams} \author[Xu et al.] {D. D. Xu$^{1,2}$\thanks{E-mail:
dandanxu@jb.man.ac.uk; dandanxu@bao.ac.cn}, Shude Mao$^{1}$, Andrew P.
Cooper$^{3}$, Jie Wang$^{3}$, Liang Gao$^{2,3}$, \and Carlos
S. Frenk$^{3}$, V. Springel$^{4,5}$ \\ $^{1}$ Jodrell Bank Centre for
Astrophysics, the University of Manchester, Alan Turing Building,
Manchester M13 9PL, United Kingdom \\ $^{2}$ National Astronomical
Observatories, Chinese Academy of Sciences, Beijing, 100012, China \\
$^{3}$ Institute for Computational Cosmology, Dept. of Physics,
University of Durham, South Road, Durham DH1 3LE, United Kingdom \\
$^{4}$ Max-Planck Institut F\"ur Astrophysik,
Karl-Schwarzshild-Stra{\ss}e 1, 85740 Garching, Germany \\ $^{5}$
Heidelberg Institute for Theoretical Studies, University of
Heidelberg, Schloss-Wolfsbrunnenweg 35, D-69118 Heidelberg, Germany }

\begin{document}
\date{Accepted ...... Received ...... ; in original form......   }

\pagerange{\pageref{firstpage}--\pageref{lastpage}} \pubyear{2009}
\maketitle
\label{firstpage}
\begin{abstract}
Lensing flux-ratio anomalies have been frequently observed and taken
as evidence for the presence of abundant dark matter substructures
in lensing galaxies, as predicted by the cold dark matter (CDM)
model of cosmogony. In previous work, we examined the cusp-caustic
relations of the multiple images of background quasars lensed by
galaxy-scale dark matter haloes, using a suite of high-resolution
$N$-body simulations (the Aquarius simulations). In this work, we
extend our previous calculations to incorporate both the baryonic
and diffuse dark components in lensing haloes. We include in each
lensing simulation: (1) a satellite galaxy population derived from a
semi-analytic model applied to the Aquarius haloes, (2) an empirical
Milky-Way globular cluster population and (3) satellite streams
(diffuse dark component) identified in the simulations. Accounting
for these extra components, we confirm our earlier conclusion that
the abundance of intrinsic substructures (dark or bright, bound or
diffuse) is not sufficient to explain the observed frequency of
cusp-caustic violations in the CLASS survey. We conclude that the
observed effect could be the result of the small number statistics
of CLASS, or intergalactic haloes along the line of sight acting as
additional sources of lensing flux anomalies. Another possibility is
that this discrepancy signals a failure of the CDM model.

\end{abstract}

\begin{keywords}
Gravitational lensing - dark matter - galaxies: ellipticals -
galaxies: formation
\end{keywords}

\section{INTRODUCTION}

The cold dark matter (CDM) cosmogony predicts that cosmic structures
form hierarchically through a succession of mergers and accretions.
The mass of the Milky Way's own dark halo is currently constrained
to $\sim1-5\times10^{12}\,M_{\sun}$ (e.g. \citealt{Guo09arxiv}).
Numerical simulations predict a wealth of dark matter substructures
(self-bound subhaloes) surviving in haloes of this mass (e.g.
\citealt{Gao2004a}; \citealt{Gao2004b, DKM08Nature,
  volker08Aq}). These substructures have a power-law mass function;
scaling this to a satellite galaxy luminosity function (LF) by
adopting a fixed mass-to-light ratio consistent with the brightest
objects overpredicts the number of satellite galaxies of the Milky
Way (MW) and M31 by a factor of several hundred $-$ the so-called
`missing satellite' problem (e.g. \citealt{Klypin1999apj,
Moore1999apj}; for a recent review see
\citealt{Kravtsov2010Review}).
For many years, theoretical models of galaxy formation have
predicted that the star formation efficiency in low-mass haloes can
be strongly suppressed through a combination of photoionization
(\citealt{Efstathiou1992}) and supernova feedback
(\citealt{WhiteRees1978}). This suppression renders many such haloes
permanently `dark' and may solve the apparent discrepancy (e.g.
\citealt{Kauffmann1993, Bullock2000, Gnedin2000, Benson02sats,
  Okamoto09Frenk, Maccio2009ExplainLF};
  \citealt{LiHelmiLucia2010}; \citealt{Stringer09arxiv}).
  Recent discoveries of low luminosity satellites
in the MW and M31 and corrections for incompleteness (e.g.
\citealt{Koposov2008MWLF, Tollerud2008}) imply reasonable agreement
between observations and theoretical predictions for MW-like haloes
(\citealt{Benson02sats}). However, because they rely on uncertain
baryonic physics and currently limited data, these comparisons so
far only provide an indirect and uncertain test of the abundance and
mass function of CDM substructures. Since gravitational lensing
traces mass, regardless of its luminosity, it can provide a powerful
alternative tool to investigate otherwise `hidden' subhaloes and
directly test the CDM model (e.g. \citealt{MKA09SubImageAnomaly,
  VKBTG2009}).

Flux-ratio anomalies observed in multiple images of lensed quasars are
often cited as evidence for substructures in lensing galaxies
(e.g. \citealt{MS1998mn, MM2001, MZ2002, DK2002, Chiba2002, KD2004,
  Metcalf2004, Chiba2005, Sugai2007, McKean2007, More2009}). Some
studies, e.g. \citet{DK2002}, \citet{BS2004aa} using CDM simulations
have concluded that the substructure population expected for a
galaxy-scale lens can reproduce the observed flux anomaly cases.
However, other studies, e.g. \citet{MaoJing04apj}, \citet{AB06mn},
\citet{Maccio2006b}, \citet{Maccio2006} and our own previous work
(\citealt{Dandan09AquI}) have argued that the simulated CDM
substructure population is \textit{not} sufficient to explain the
observed frequency of lensing flux-ratio anomalies. This negative
conclusion does not yet rule out the predictions of CDM, as the
observed sample of `anomalous' lenses is small. Intergalactic haloes
along the line of sight may also perturb the lensing potential and
cause the observed lensing anomalies (\citealt{Chen2003,
  Wambsganss2005, Metcalf2005a, Metcalf2005b}; \citealt{Miranda2007,
  PunchweinHilbert2009}).

One extreme case of flux-ratio anomalies is the cusp-caustic
violation,
which has been observed in five lens systems of a cusp geometry:
B0712+472 (\citealt{Jackson2000B0712,Jackson98B0712}), B1422+231
(\citealt{Impey1996B1422,Patnaik2001B1422}), B2045+265
(\citealt{Fassnacht1999B2045}), RXJ1131-1231
(\citealt{Sluse2003aaJ1131}) and RXJ0911+0551
(\citealt{Bade1997aa317, Burud1998apjl}). In the first three cases
(those detected in radio observations) these have been attributed to
substructure lensing, while the remaining two are thought to be due
to microlensing (\citealt{Anguita2008Microlensing,
Morgan2006RXJ1131}).  In \citet{Dandan09AquI} (which we summarise
below) we found that even with the well-resolved subhalo population
from the Aquarius simulations, the observed cusp lenses still
violate the cusp-caustic relation more frequently than expected for
a concordance CDM model. However, our earlier work did not consider
the effect of baryons concentrated at the centre of subhaloes, which
may modify their density profiles and so alter their lensing
properties. In this work, we carry out a more thorough analysis of
the lensing effect of baryons in substructures, proceeding as
follows. In Section 2 we use a semi-analytic model of galaxy
formation to predict the baryon content of subhaloes in each
Aquarius simulation: we model the density distribution of baryons in
each subhalo and re-run our lensing simulations to study
cusp-caustic violations. In Section 3, to estimate the effects of
possible baryonic concentrations below the resolution limit of the
simulation, we further include an empirical Galactic globular
cluster population (Harris 1996) in our model. Finally, as typical
surface density fluctuations of only a few per cent would be
sufficient to cause flux-ratio anomalies (\citealt{MS1998mn,
Li2006SPH}), we consider irregularities in the surface density of
the main halo due to the debris of tidally disrupted satellites (the
diffuse component) in Section 4. Section 5 provides a short summary
and a brief discussion of our results.

\section{Semi-analytic Galaxies and their Cusp-caustic Violations}
\label{sec:GalaxyCatalogue}
\begin{figure*}
\centering
\includegraphics[width=16cm]{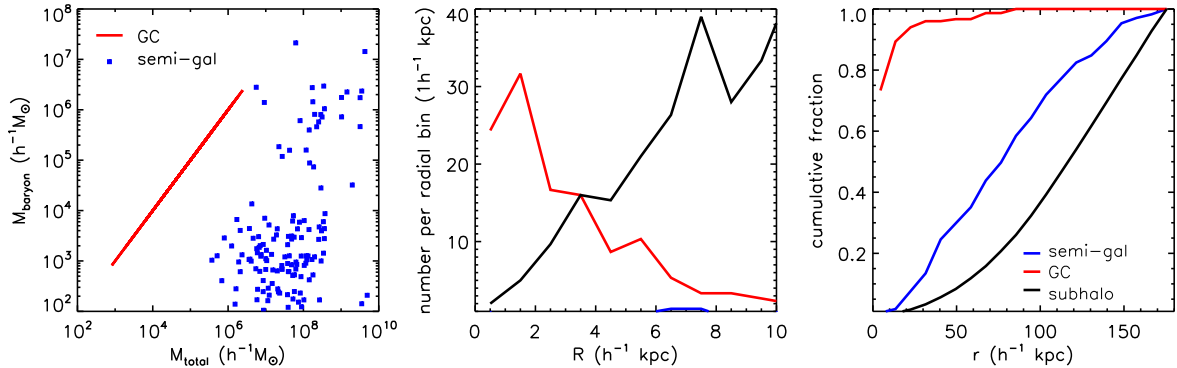}
\caption{The masses and radial distributions of semi-analytic galaxies
  (blue) and dark matter subhaloes (black) from {\it Aq-A-2}, as well
  as the observed Milky-Way globular clusters (red; using the
  catalogue from \citealt{HarrisGC1996}). The panel on the left shows
  the stellar mass $M_{\rm baryon}$ as a function of the total mass
  $M_{\rm total}$, which in the case of semi-analytic galaxies is the
  total mass of a satellite galaxy and its host subhalo. The dynamical
  masses of globular clusters come from their observed V-band
  luminosities, assuming a mass-to-light ratio to be 3. Such masses
  are equal to their stellar masses $M_{\rm baryon}$, assuming a zero
  dark matter content. In general, globular clusters have lower
  dynamical masses than satellite galaxies, but the latter have higher
  mass-to-light ratios than the former. The middle and right panels
  present the projected and 3D radial distributions of the three
  components. Globular clusters are much more concentrated towards the
  galactic centre than satellite galaxies. At the Einstein radius of
  the main halo ($2 \sim 4 h^{-1}$ kpc), the number of projected
  globular clusters is about $1 \sim 2$ times that of dark matter
  subhaloes. Satellite galaxies are rare in the projected central
  region.} \label{fig:C02type12GpropertyCompare}
\end{figure*}

\subsection{The Aquarius Simulations}

The Aquarius project (\citealt{springel08_nature};
\citealt{volker08Aq}) is a suite of collisionless $\Lambda$CDM
$N$-body simulations run with the {\sc gadget-3} code. Each of the six
simulations in the suite (labelled A-F) centres on a single halo of
mass $\sim10^{12}h^{-1}M_{\sun}$. These haloes were selected for
high-resolution multi-mass resimulation from a
100$h^{-1}\,\rm{Mpc^{3}}$ simulation with a uniform mass resolution: a
subsequent high-resolution study of this parent volume (the Millennium
II simulation) confirms that in almost all respects the Aquarius
sample is representative of haloes of this mass
(\citealt{Millennium2}; \citealt{BoylanKolchin09arxiv}).

All six haloes were simulated at a resolution corresponding to a
particle mass of $\sim10^{4}h^{-1}M_{\sun}$ and a softening length
of $\sim 50 h^{-1}$ pc, resulting in $\sim130$ million particles
within the virial radius of the main halo at $z=0$. This resolution
of the Aquarius project is referred to as `level 2' and the six
haloes at this resolution are labelled \textit{Aq-A-2, Aq-B-2,
Aq-C-2, Aq-D-2, Aq-E-2} and \textit{Aq-F-2}. At this resolution,
subhaloes are resolved to a mass limit of $ \msub \ga 10^5
h^{-1}M_{\odot}$ (corresponding to bound associations of 20
particles identified by the {\sc subfind} algorithm). We use these
level 2 simulations in this work. Simulations of one halo (Aq-A) at
higher and lower
  resolutions have demonstrated that the abundance, radial
  distribution and internal structure of subhaloes relevant to our
  lensing calculations are numerically converged to high accuracy at
  the resolution of the level 2 simulations (full details of these
  extensive convergence studies are given by \citet{volker08Aq} and
  \citet{Navarro10aq}).

Throughout this work, the cosmology of our lensing simulations is
the same as that used for the Aquarius project, with a matter
density $\Omega_{\rm m}$ = 0.25, cosmological constant
$\Omega_{\Lambda}$ = 0.75, Hubble constant
$h=H_0/(100\kms\Mpc^{-1})=0.73$ and linear fluctuation amplitude
$\sigma_8=0.9$.

\subsection{Halo lensing model}

In \citet{Dandan09AquI}, we studied cusp-caustic violation
  probabilities using the six galaxy-scale CDM haloes provided by the
Aquarius simulations. Mock lensing signals from background quasars
were used to determine the observed frequency of flux-ratio anomalies
due to dark matter substructures in these haloes. The mass resolution
of Aquarius is at least two orders of magnitude higher than that of
previous simulations used for lensing studies, resolving bound
subhaloes down to a mass of $\sim10^5 h^{-1}M_{\sun}$. In addition to
the simulated dark matter distribution, we added a Hernquist profile
`galaxy' at the centre of each main halo. We renormalized the dark
matter particle mass and adiabatically contracted the particle
distribution to account for this extra contribution to the potential
(e.g.  \citealt{BarnesWhite1984, Blumenthal1986AC, MoMaoWhite1998,
  Gnedin2004}).

A particle-mesh (PM) code was applied to calculate lensing
potentials, deflection angles and magnifications from the simulated
lensing haloes. The numerical accuracy of this code is described
fully in \citet{Dandan09AquI}. Near the critical curves, high
magnification makes changes in the image properties sensitive to
small perturbations in surface density. To estimate the effects of
numerical noise in these regions, we used our code to calculate the
lensing properties of isothermal ellipsoids generated by Monte Carlo
sampling, using an equivalent number of particles to the level 2
Aquarius simulations ($\sim 10^8$ particles within the radius
$r_{200}$\footnote{Defined as the radius enclosing a mean density of
 200 times the critical density of the Universe.}).
 The numerical accuracy of the deflection angle, convergence (surface
density) and magnification for these Monte Carlo realisations are
shown in Figure 4 of \citet{Dandan09AquI}. The uncertainties around
the Einstein radius (at about 0.02 $r_{200}$) are 0.2\% for the
deflection angle, 1\% for the convergence, and $<$ 10\% for the
  magnification. Figure 6 of \citet{Dandan09AquI} presents the
  probability distribution function of the image flux ratio $\Rcusp$
  (Eq. \ref{eq:Rcusp}) for cusp sources with an image opening angle
  smaller than 90$^{\circ}$ in the $10^8$-particle Monte Carlo
  realisations. This distribution was found to be broader than the
  equivalent analytical calculation as the result of discreteness
  noise, inherent in mapping an $N$-body density field to a mesh.

To correct for this sampling noise, each main Aquarius halo
(including a central `galaxy') was modelled as an isothermal
ellipsoid (e.g. \citealt{Rusin2003, RusinKochanek2005,
Koopmans2006apj,
  Gavazzi2007}), the lensing properties of which can be solved
analytically. These ellipsoids were determined as best fits to the
critical curves and caustics of the main galaxy haloes. The
substructure population of each halo was taken directly from the
simulations and superimposed on the analytic density fields of these
isothermal ellipsoids. Multiple images of mock background quasars
were recovered through an image-finding procedure combining the
Newton-Raphson and Triangulation methods (for a more detailed
description of this procedure, see \citet{Dandan09AquI}). The
  cusp-caustic relations and their violation probabilities were
  calculated for the simulated lensing-galaxy haloes and compared to
  those derived from observations. In this paper we extend this method
  to study additional contributions to the lensing potential.

\subsection{Semi-analytic galaxies}

The Aquarius simulations provide us with particle distributions of
the main dark matter haloes and an associated population of
subhaloes. To investigate the effect of baryons in substructures, we
postprocess Aquarius with the {\sc galform} semi-analytic model
(\citealt{Cole1994, Cole2000, Bower2006}). We determine the
  dark matter merger history of each subhalo directly from the
  simulations \citep[][]{Helly03}. From each of these `merger trees',
  {\sc galform} computes the history of gas accretion, star formation
  and feedback in the corresponding galaxy. Here, as a fiducial model,
  we adopt the parameter values of \citet{Bower2006}, which have been
  shown to be consistent with a number of observational
  constraints. As discussed by \citet{CooperStream2009}, the
\citet{Bower2006} parameters also agree well with the MW and M31
satellite galaxy LFs, provided that the value of the parameter $V_{\rm
  cut}$ is set to $\sim30\,\mathrm{km\,s^{{-}1}}$ (a value consistent
with recent estimates from numerical simulations\footnote{$V_{\rm
    cut}$ is defined as the halo circular velocity below which a
  photoionizing background strongly suppresses further baryon
  accretion and cooling. Reionization is assumed to occur
  instantaneously at $z=6$ in the \citet{Bower2006} model.}), in
preference to the value of $50\,\mathrm{km\,s^{{-}1}}$ used by
\citet{Bower2006}. A discrepancy with observations is only apparent
at $M_{V}\ga -5$, where constraints on the observed LF are weak. We
stress that {\sc galform} is only used here to determine which dark
matter substructures host stars, and to assign reasonable
mass-to-light ratios to those substructures. For this purpose, a
fiducial model producing a satellite LF consistent with the
available data is sufficient. Our choice of the \citet{Bower2006}
parameters with $V_{\rm cut}=30\,\mathrm{km\,s^{{-}1}}$ is identical
to that used by \citet{CooperStream2009} as a fiducial model in
their study of galactic stellar haloes. Our findings concerning the
lensing effect of substructures are not strongly sensitive to
further changes in this model that could be made to match other
properties of the satellite population.

Populating the resolved subhaloes of Aquarius with our fiducial {\sc
  galform} model results in $\sim 100-200$ satellite galaxies within
the virial radius of each Aquarius halo. As an example, Fig.
\ref{fig:C02type12GpropertyCompare} presents the properties of the
semi-analytic galaxies (shown as blue symbols) in the halo {\it
  Aq-A-2}. The combined mass, $M_{\rm total}$, of a satellite galaxy
and its host subhalo ranges from $\sim 10^6 h^{-1} M_{\odot}$ to
$\sim 10^{10} h^{-1} M_{\odot}$ with a median of $ \sim 3 \times
10^7 h^{-1} M_{\odot}$. The baryonic mass $M_{\rm baryon}$ spans
more than six decades, which is consistent with the wide range of MW
satellite luminosities (from $\sim 100 L_{\odot}$ to $\sim 10^8
L_{\odot}$, e.g. \citealt{Strigari2008Nature, Koposov2008MWLF,
Stringer09arxiv}). Their radial distribution has a median
halo-centric distance of $70 h^{-1}$ kpc. It is rare to find a
satellite galaxy projected within the Einstein radius of the main
halo ($2 \sim 4 h^{-1}$ kpc).

We require a structural model for each satellite galaxy in its host
subhalo. Although {\sc galform} predicts the morphology and size of
each simulated galaxy, we prefer to adopt a simpler and more
conservative model that \textit{maximises} the lensing effect. We
describe all satellite galaxies using a singular isothermal sphere
(SIS) density profile. This profile, of the form $\rho(r) \propto
r^{-2}$, leads to a constant deflection angle. The SIS profile of each
galaxy terminates at a radius $r_{\rm t}$, which we set to be the
half-mass radius $r_{\rm H}$ of the galaxy's host dark matter subhalo
(resulting in a median truncation radius of $\sim 200\,h^{-1}$
pc). The total stellar mass of the galaxy predicted by {\sc galform}
is enclosed within $r_{\rm t}$. Under these assumptions, the median
Einstein radius is $\sim 0.006^{\prime\prime}$. Further increasing the
concentration of our stellar components by setting $r_{\rm t} =
0.3\,r_{\rm H}$ (median $r_{\rm t} \sim 60\,h^{-1}$ pc) makes little
difference to the final cusp-caustic violation probability (as
discussed below, this is sensitive to the spatial distribution of the
substructures in the main halo - see
Fig. \ref{fig:C02type12GpropertyCompare}). All satellites have high
mass-to-light ratios such that adiabatic contraction due to the
baryonic component can be neglected. The full hierarchy of
sub-subhaloes within subhaloes is included in the lensing calculation,
as in our earlier dark matter-only study. Galaxies hosted by these
sub-subhaloes are also represented by SIS profiles in this updated
calculation.

To determine the new cusp-caustic violations that result from
including baryons in dark substructures, the semi-analytic galaxies
are added to the dark matter density fields from Aquarius. For the
dark matter component of each subhalo, the particle distribution
from the simulation is used directly, as in our earlier study; we
only assume SIS profiles for the stars. The new density fields are
processed through our PM code and the image-finding routine to
recreate lensing signals of background quasars.

\subsection{The cusp-caustic violation}

In any smooth lensing potential an asymptotic magnification relation
(the `cusp-caustic relation'; \citealt{BN1986apj};
\citealt{SW1992aa}; \citealt{Zakharov1995AA}; \citealt{KGP2003apj})
will hold:
\begin{equation}
  \Rcusp \equiv \frac{|\mu_A + \mu_B +
    \mu_C|}{|\mu_A|+|\mu_B|+|\mu_C|} \rightarrow 0,
\label{eq:Rcusp}
\end{equation}
where $\mu$ denotes the magnifications of the three closest images
$A$, $B$ and $C$ of a background source, which is located near a
cusp of the central caustic. In this case, the total absolute
magnification $|\mu_{\rm A}|+|\mu_{\rm B}|+|\mu_{\rm C}|$ of the
triple images goes to infinity. Following the same procedure as
\citet{Dandan09AquI}, the lensing haloes are located at redshift 0.6
and the background quasars at redshift 3.0. `Cusp sources' are
defined as sources (quasars) that are near the cusps of the central
caustic and have image opening angles, $\Delta\theta$ of their close
triple images smaller than 90$^{\circ}$. Any cusp source with
$R_{\rm cusp} \geqslant 0.187$ is defined as a cusp-violation case,
as 0.187 is the observed value for B1422+231, which shows the
smallest violation (smallest $\Rcusp$ value) among the five observed
cusp lenses. In each of the three independent projections from each
Aquarius halo, approximately 10000 - 20000 cusp sources are
generated and used to calculate the $\Rcusp$ distribution and the
violation probability.

\begin{table*}
\centering
\caption{The average substructure surface mass fractions and the
  cusp-caustic violation probabilities for scenarios in which
  additional contributions to the lensing potential are
  considered. Col. 2: $f_{\rm sub, annu}$ is the mean surface mass
  fraction of substructures within an $0.1^{\prime\prime}$-annulus
  around the tangential critical curve. Col. 3: $P(R_{\rm
  cusp}\geqslant 0.187)$ is the mean cusp-violation rate averaged over
  all projections from six haloes, under which Col. 4, $P_{3|5}$,
  gives the probability of observing 3 out of 5 cusp lenses violating
  the cusp-caustic relation due to the presence of substructures.}
  \small\addtolength{\tabcolsep}{-1.5pt}
\label{tab:G12LensingViolation}
\begin{minipage} {\textwidth}
\begin{tabular}[b]{l|c|c|c}\hline
Case & $~~f_{\rm sub, annu}~ $ & $~~ P(R_{\rm cusp}\geqslant
0.187)~$ & $~~P_{3|5}~~$ \\ \hline (a), dark matter subhaloes only &
0.20\% & 10.1\% & 0.8\% \\ (b), (a) $+$ semi-analytic galaxies &
0.21\% & 10.2\% & 0.9\% \\ (c), (b) $+$ Milky-Way globular clusters
& 0.23\% & 11.0\% & 1.1\% \\ \hline
\end{tabular}
\end{minipage}
\end{table*}

\begin{figure*}
\includegraphics[width=8cm]{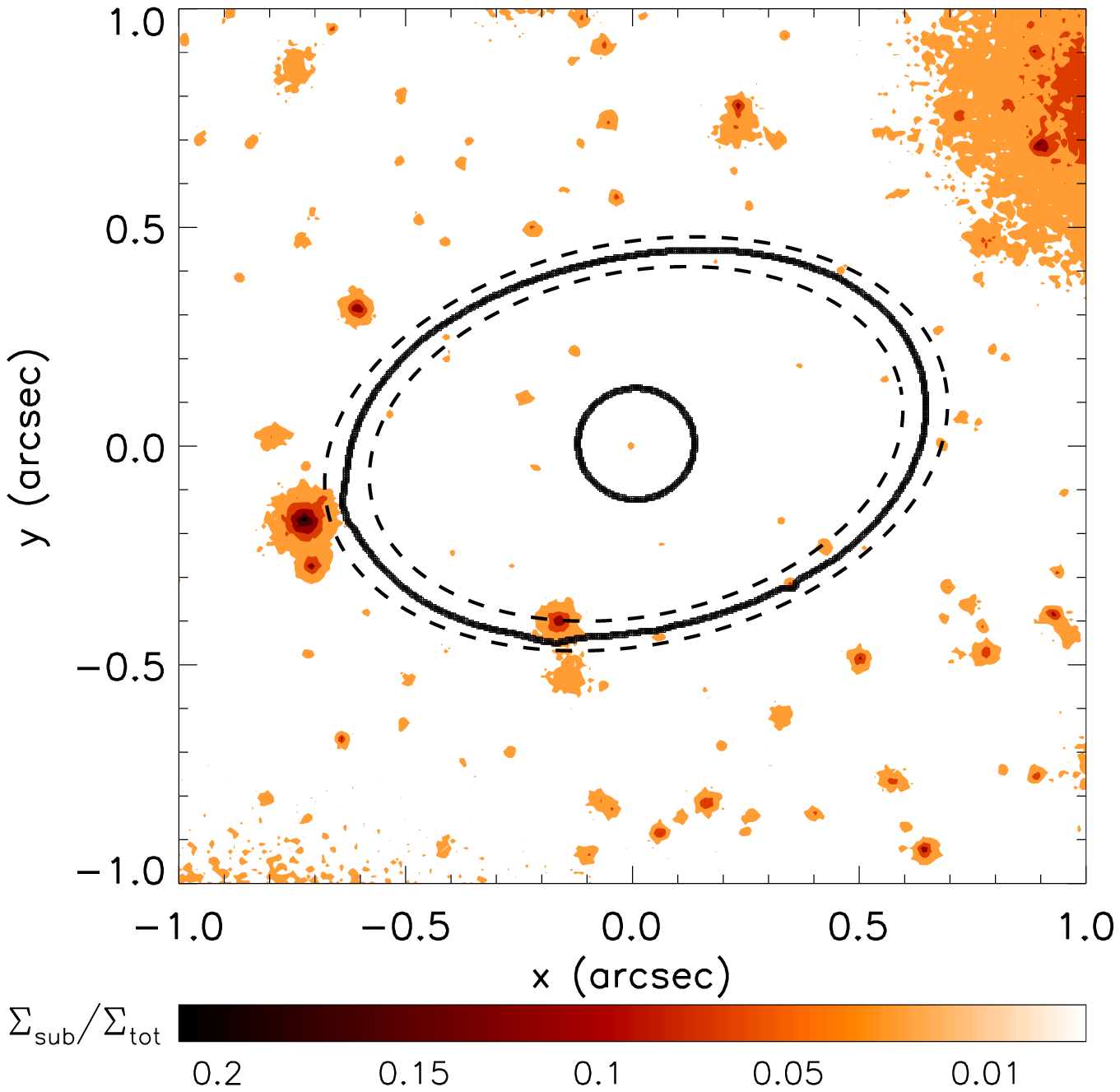}
\includegraphics[width=8cm]{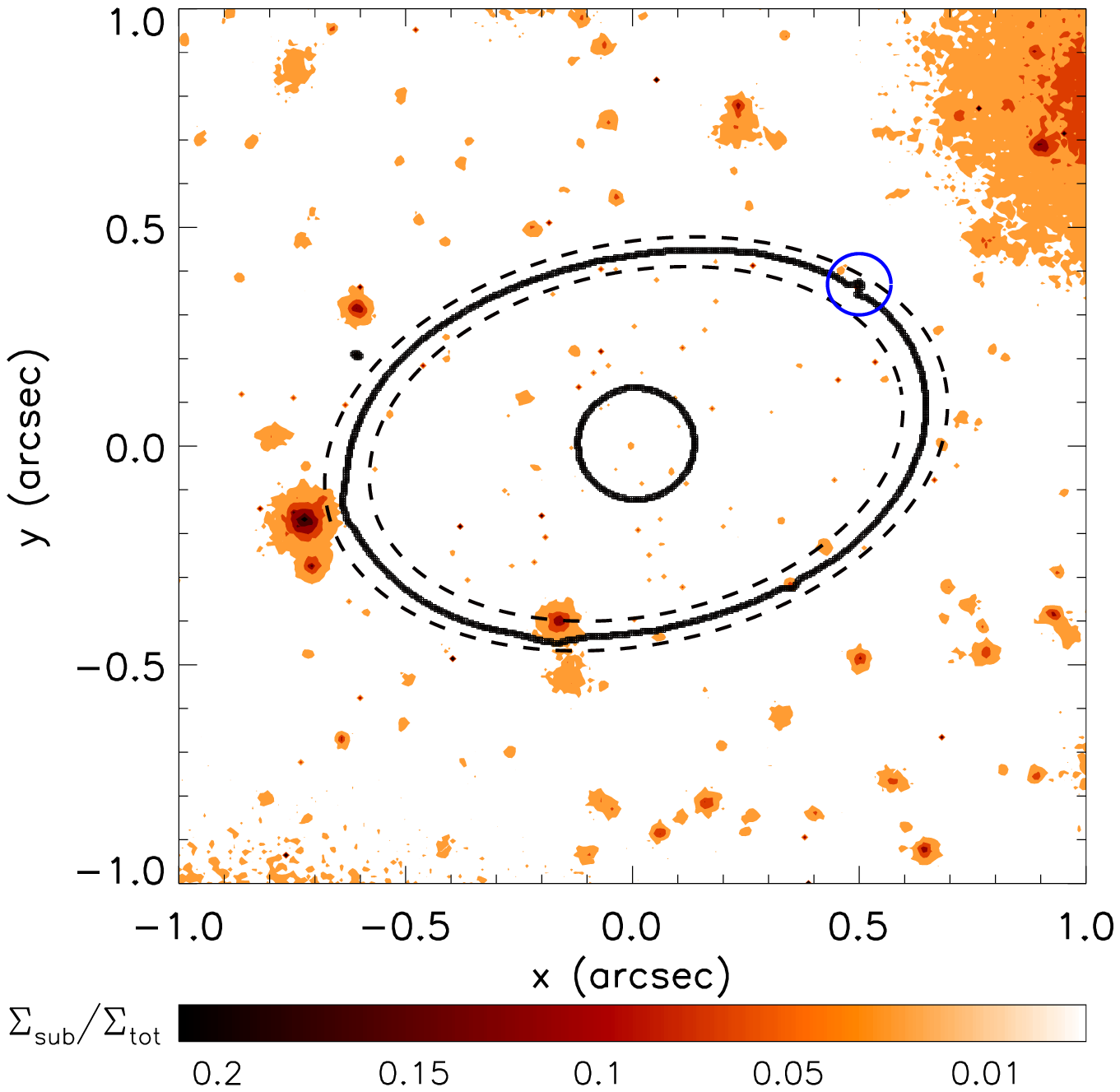}
\caption{Contour maps of the substructure surface mass fraction of the
  halo {\it Aq-E-2}, in $X$-projection. Left: semi-analytic galaxies
  are added to the dark matter subhalo population. The mean surface
  mass fraction, $f_{\rm sub, annu}$ of the substructures within the
  $0.1^{\prime\prime}$-annulus (indicated by the dashed lines) around
  the tangential critical curve is $\sim$0.16\%. Right: semi-analytic
  galaxies and Milky-Way globular clusters are added to the dark
  subhalo population. $f_{\rm sub, annu} \sim 0.19\%$. Globular
  clusters are more centrally distributed in the projected central
  region than satellite galaxies. The blue circle indicates an example
  of the small-scale wiggles induced by globular clusters.}
\label{fig:C02AddType1}
\end{figure*}

Our earlier work (\citealt{Dandan09AquI}) found an average
cusp-violation rate of $\sim 10\%$ over all three independent
projections of each of the six haloes\footnote{There were four
  projections in total with naked cusps (see \citealt{Dandan09AquI})
  present in the central caustic due to the large core sizes
  of the imposed Hernquist profiles. The exclusion of these
  four cases in the final calculation resulted in a mean
  cusp-violation rate of 6.4\%. Here in order to avoid these ``naked
  cusps'', the core size is artificially set to be
  0.05$^{\prime\prime}$ for all six haloes, leading to the higher
  cusp-violation rate of $\sim 10\%$ as shown in Table
  \ref{tab:G12LensingViolation} case (a).}, which leads to a
probability $< 0.01$ of observing 3 out of 5 cusp lenses violating
the cusp-caustic relation due to substructures. This result is
listed as case (a) in Table \ref{tab:G12LensingViolation}, which
summarises the effects of additional contributions to the lensing
potential.

Case (b) in Table \ref{tab:G12LensingViolation} adds the
semi-analytic galaxies described above to the dark matter
substructures. The change in the cusp-caustic violation is small.
The reason for this is clear from Fig. \ref{fig:C02AddType1} (left),
which shows the critical curves superimposed on a contour map of the
substructure surface mass fraction, in the halo {\it Aq-E-2} (a
single projection is shown as an example; our results are averaged
over many projections). This figure can be compared with figure 11
in \citet{Dandan09AquI}: the addition of satellite galaxies alters
the density profile of individual subhaloes, but cannot alter the
abundance of substructures around the tangential critical curve,
where the cusp-caustic relation is examined. Therefore, the baryonic
`boost' from the small number of galaxy-hosting subhaloes
contributing to the overall cusp-caustic violations is marginal.

As shown in Table 3 of \citet{Dandan09AquI}, the violation
probability varies from halo to halo, and from projection to
  projection within the same halo. The lensing cross-section for
  cusp-caustic violation is significant only where massive
  substructures are projected onto the central region. The presence
  or absence of these strong perturbations to the lensing potential is
  highly stochastic, because massive subhaloes are intrinsically rare
  and are typically found in the outer regions of their hosts. It is
  therefore possible in principle that a limited sample of haloes and
  projections could lead to an underestimate (or overestimate) of the
  violation probability. However, the mass function and radial
  distribution of subhaloes is found to be very similar among all six
  Aquarius haloes \citep{volker08Aq}. In these respects the Aquarius
  haloes are themselves representative of the population from which
  they were selected \citep{BoylanKolchin09arxiv}. Our sample of 18
  projections (combining three projections in six similar haloes) is
  therefore unlikely to significantly underestimate the violation
  probability.

\section{Milky-Way Globular Clusters and Their Cusp-caustic Violations}

Scenarios for the formation of globular clusters (GCs) are numerous
and highly uncertain. GCs are known to have low mass-to-light ratios
(i.e. to be baryon-dominated) at the present day
(e.g. \citealt{MarastonGCML2005}), although a past association between
individual GCs and dark matter haloes is not strongly ruled out. It
may be that some GCs are the relics of the very first generations of
star formation in the early Universe (e.g.  \citealt{Gao2009,
  Griffen2009}). It is often speculated that a number of GCs (most
notably $\omega$ Centauri) should be identified with the stripped
nuclei of dwarf satellite galaxies. This suggests that the definitions
of these two classes of object might be ambiguous.  At the present day
GCs are strongly concentrated around central galaxies (more so than
the overall subhalo populations in simulated CDM haloes) and their
survival is known to be subject to many factors, including evaporation
and tidal disruption. GCs are likely to be present in lensing galaxies
in large numbers and may perturb the gravitational potentials in the
inner regions, causing cusp-caustic violations. A simple estimate of
their contribution was made by \citet{MS1998mn}, who concluded that a
surface density fluctuation of a few per cent ($\delta\kappa\sim
0.01$) from GCs would be enough to cause the observed flux-ratio
anomaly in B1422+231. This conclusion has been re-examined in this
work.

Here we adopt an empirical approach to the effects of GCs on the
lensing potential. We use the catalogue of Milky-Way GCs from
\citet{HarrisGC1996}, which provides their spatial distribution,
V-band luminosities $L_{\rm v}$ and half-mass radii $r_{\rm
  h}$. Although the Milky-Way GC distribution is slightly flattened within
the central $\sim 10 h^{-1}$ kpc, the choice of projection does not
affect our results.

It is interesting to ask whether a proportion of the `satellite
galaxies' in Aquarius should in fact be identified with a population
of `primordial' galaxy-like objects or with the `cores' of galaxies
that have been heavily stripped. We have already included satellites
in our calculation, so our approach to GCs risks double-counting
some objects if either of these cases is true. However, a detailed
investigation of this issue is beyond the scope of this paper, and
we will assume that most GCs are not already represented as
`satellites' in our semi-analytic model. As we state above, the LF
of bright satellite galaxies in our model matches the shape of the
Milky-Way satellite LF. This observed LF does not include the many
equally bright but structurally distinct Milky-Way GCs: this in turn
suggests that these bright GCs are not represented by some of the
existing `satellites' in our model, stripped or otherwise. Fainter
than $M_{V}\sim-5$ the distinction between galaxies and clusters is
much less certain and the interpretation of the current data is not
at all clear. However, these low-mass objects are not significant
for lensing.

The masses of our empirical Milky-Way GCs are obtained assuming
$M/L_{\rm v} = 3$ (e.g. \citealt{MarastonGCML2005}). We adopt a SIS
profile for each GC to maximise our estimate of its lensing effect. We
choose the SIS truncation radius $r_{\rm t} = 2 r_{\rm h}$ such that
the model half-mass radius is equal to the GC half-light radius
($M_{\rm SIS} (<r) \propto r$). The median Einstein radius is $\sim
0.003^{\prime\prime}$ under these assumptions. Fig.
\ref{fig:C02type12GpropertyCompare} summarises our empirical GC
population, showing the masses and radial distributions of the
Milky-Way GCs (red symbols). Their dynamical masses range from
hundreds to millions of solar masses (lower than that of the majority
of satellite galaxies). The radial distribution of GCs is much more
concentrated than satellite galaxies: the number of GCs projected at
the Einstein radius of the main halo is about $1 \sim 2$ times the
number of the dark matter subhaloes at that radius.

Comparing the right and left panels in Fig. \ref{fig:C02AddType1},
more substructures (GCs) in this case are projected in the vicinity
of the tangential critical curves, inducing more smaller-scale
wiggles; one example is indicated by the blue circle. Case (c) in
Table \ref{tab:G12LensingViolation} summarises our lensing results
after including the empirical GC population. The mean cusp-violation
rate averaged over all projections from six Aquarius haloes has only
marginally increased from $\sim 10\%$ to $\sim 11\%$, and the
probability to observe 3 out of 5 cusp lenses violating the
cusp-caustic relation due to substructures has increased from
$0.9\%$ to $1.1\%$. Although GCs are numerous and more concentrated
than the subhalo population, the small mass and size of a typical GC
limits their effect on the lensing potential. The predicted
violation rate is still far below that observed, so the inclusion of
GCs still does not solve the apparent lack of substructures
suggested by observations of lensing flux anomalies.

\section{Aquarius Streams and Their Cusp-caustic Violations}

\begin{figure*}
\centering
\includegraphics[width=6.5cm]{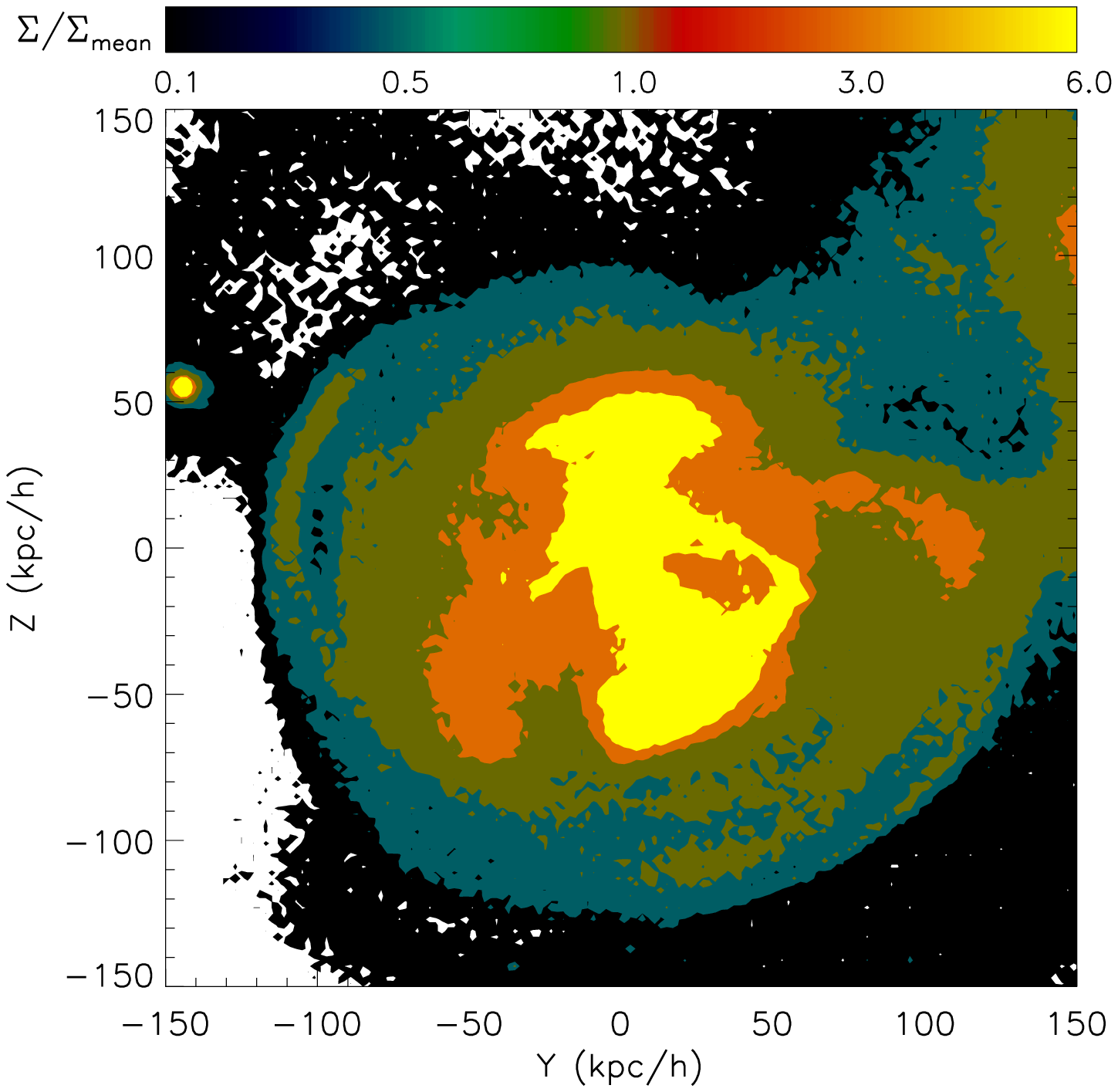}
\includegraphics[width=6.5cm]{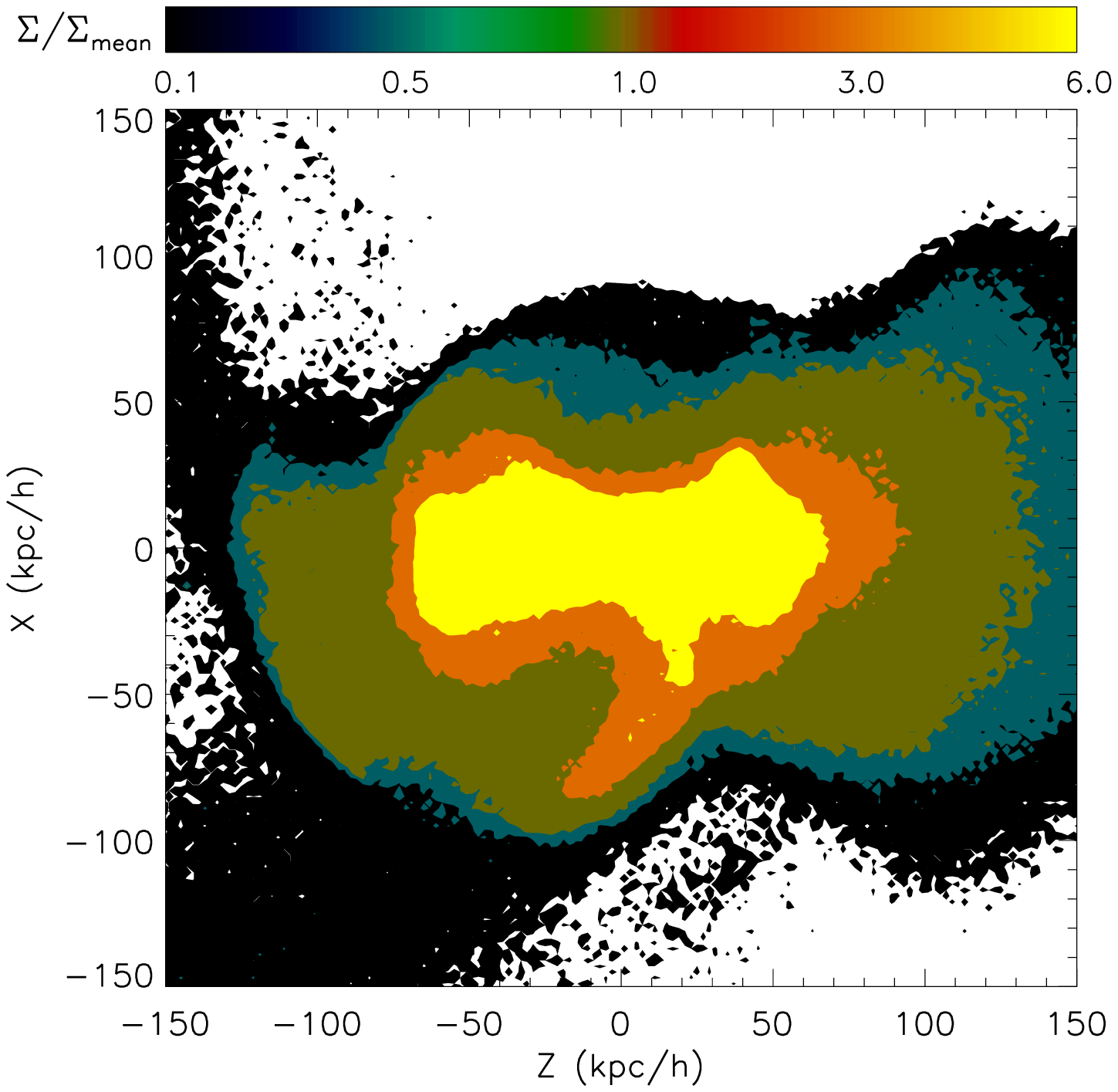} \\
\includegraphics[width=6.5cm]{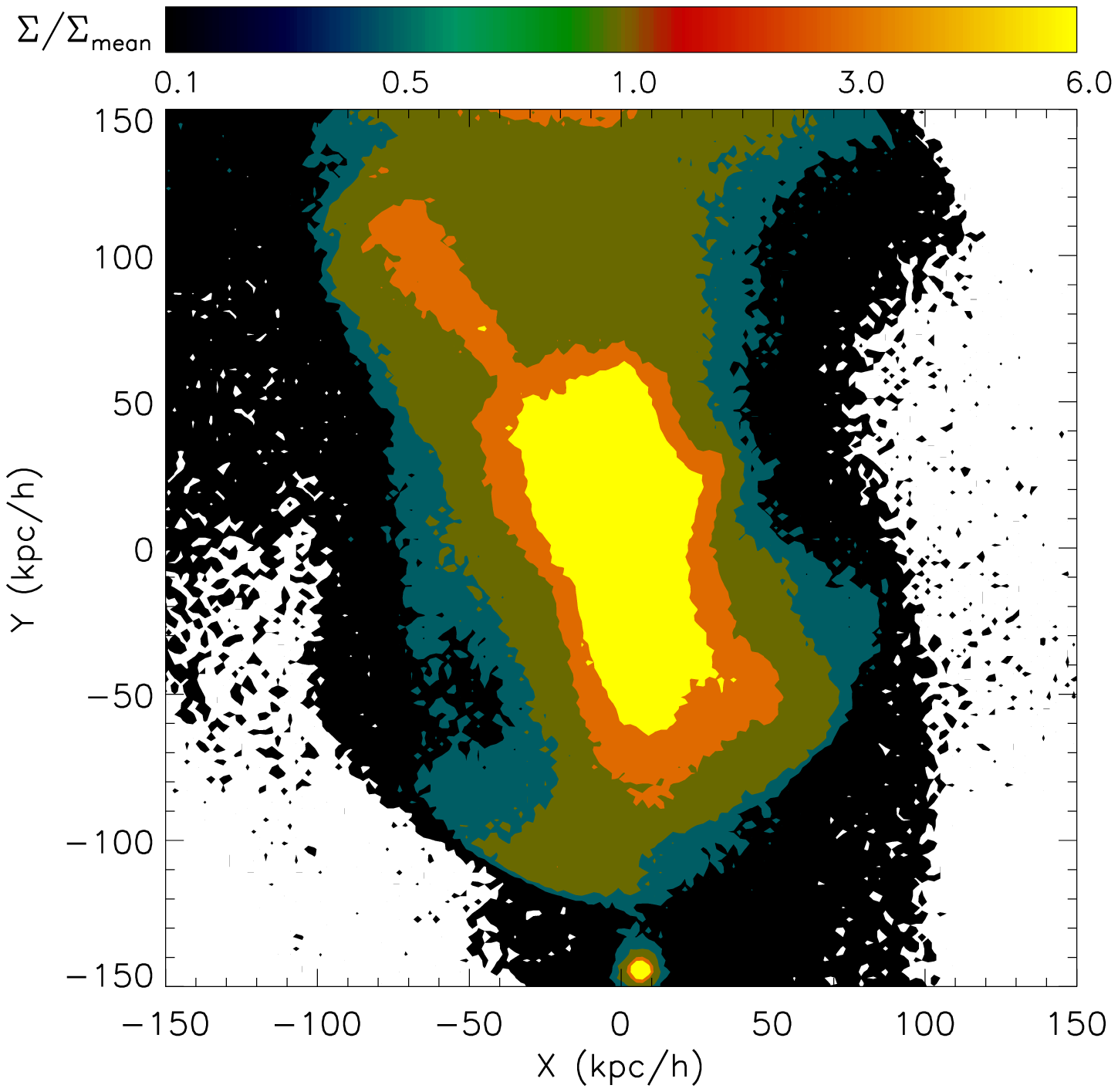}
\includegraphics[width=6.5cm]{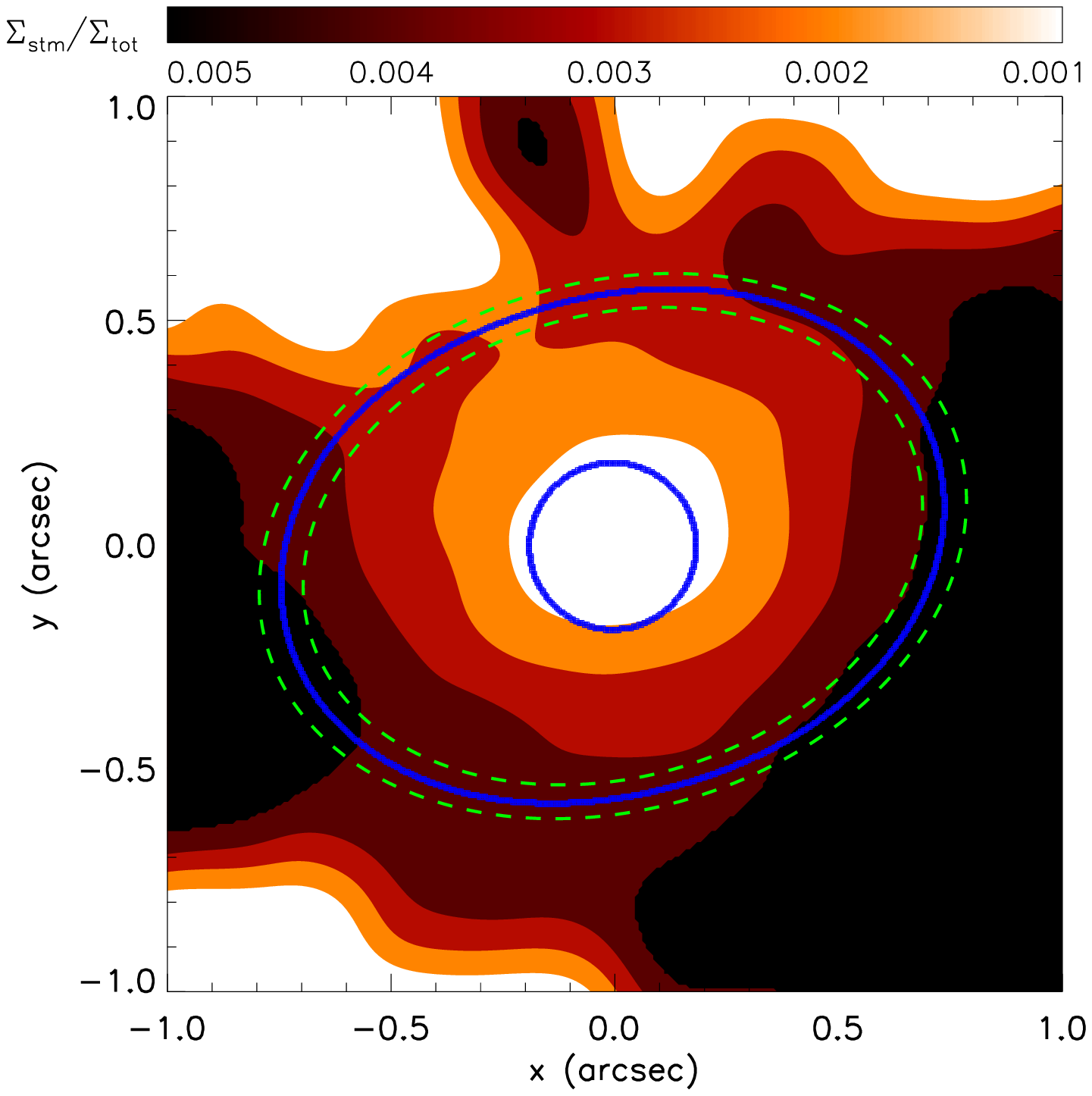}
\caption{An example of one spatially coherent stream in the
  halo {\it Aq-A-2}. The top two and bottom left panels show the
  stream on large scales in three independent projections. The bottom
  right panel presents the contour map of the surface mass density
  fraction of the stream on smaller scales in $Y$-projection,
  with critical curves overlaid; $f_{\rm stm, annu} \sim
  0.45\%$. }
\label{fig:streams1}
\end{figure*}

In recent years, streams and other stellar overdensities have been
detected in the halo of the Milky Way (e.g.
\citealt{BelokurovStream2007a}; \citealt{DiffBelokurovStream2007b})
and other nearby galaxies, most notably M31
(e.g. \citealt{McConnachie09}).  Cosmological $N$-body simulations
have demonstrated that a wealth of similar structures consistent with
these observations are produced by the tidal stripping of satellite
galaxies in typical Galactic haloes (e.g. \citealt{HelmiWhite1999,
CooperStream2009}). As stars in satellites are likely to be deeply
embedded in their own host dark matter (sub)haloes at the time of
accretion, each stellar stream in these models is naturally associated
with a more massive stream of dark matter. These `streams' exhibit a
variety of complex morphologies (e.g. Cooper et al 2009), very few of
which remain `coherent' in configuration space for long. However,
their presence might induce irregularities in the surface density of
the dark matter halo and cause lensing flux anomalies.

We make an estimate of the magnitude of this effect with the level-2
Aquarius simulations. We define streams as dark matter particles
that were once bound to substructures crossing the virial radius of
the main halo, but which at $z=0$ are bound to the main halo itself
(i.e. are no longer bound to any substructure). In the previous
sections and in \citet{Dandan09AquI}, all these particles were
treated as the smooth component of the main halo (included in the
smooth ellipsoidal fit). The streams from different infalling
substructures are phase-mixed to different degrees. Those most
strongly mixed are smoothly distributed as ellipsoids concentric
with the main halo, but a small quantity of satellite (dark matter)
debris remains in the form of coherent features in configuration
space (canonical `streams'). These features introduce the most
significant surface density irregularities.

Here we select the most massive spatially coherent streams that are
projected within the central $\sim 10 h^{-1}$ kpc region of each
halo. We do so by visually inspecting the density fields of all
massive streams ($M_{\rm DM}>5\times10^{8}h^{-1}M_{\odot}$) for
regions of high overdensity projected in the vicinity of the
critical curve. We have calculated the lensing effect of each
individual candidate stream by superimposing its full particle
distribution onto our smooth isothermal ellipsoid model for the main
halo, as we did for the bound substructures described previously. In
practice the overlap of many such streams at the centre of a lensing
galaxy decreases the density contrast of any single stream; by
isolating these streams from the smoothly distributed dark matter,
we are likely to \textit{overestimate} their overall contribution.

Fig. \ref{fig:streams1} shows one example of the massive coherent streams
meeting our criteria. The surface density contours are overlaid by the
lensing critical curves. The mean surface mass fraction, $f_{\rm stm,
  annu}$, of the stream within the $0.1^{\prime\prime}$-annulus around
the tangential critical curve, varies from $0.01\%$ to $<1\%$
depending on projection. The cusp-violation rates in all cases are
found to be extremely low, $P(R_{\rm cusp} \geqslant 0.187) \ll
1\%$, even for projections with $f_{\rm stm, annu}$ as high as those
from compact substructures. The typical width of these `dense'
spatially coherent streams is of the order of $\sim$ kpc (e.g.
\citealt{HelmiWhite1999}): this is about the size of the strong
lensing domain and hence these streams are incapable of inducing
small-scale wiggles in the tangential critical curves. On the other
hand, thin streams from lower mass progenitors are not sufficiently
dense. We conclude that the presence of dark matter streams does not
significantly boost the cusp-caustic violations.

\section{DISCUSSION AND CONCLUSIONS}

We briefly discuss a few issues which may affect our conclusions.
Firstly, the observed sample is small, consisting of only five cusp
lenses, among which are the three systems detected in the radio band
by the CLASS survey (\citealt{Myers2003, Browne2003}). This survey
shows an anomalously high incidence of secondary lenses (bright
satellites) (\citealt{BMK2008,Jackson2009}).

A larger sample of galaxy-scale lenses from future instruments will
provide better statistical constraints on substructure abundance in
galaxy-sized CDM haloes \citep{VK2009}. New lenses are
  likely to be discovered in optical surveys (such as PanSTARRS and
  LSST), although radio and mid-infrared follow-up will be necessary
  to distinguish anomalies attributable to substructure from
  microlensing events \citep[e.g.][]{WambsganssPaczynskiSchneider1990,
    WittMaoSchechter1995, Pooley2007, Eigenbrod2008}. High-resolution
  radio surveys (to be carried out by LOFAR and SKA, for example) will
  be a productive way to discover more of these systems.

Secondly, most lensing galaxies are massive elliptical galaxies due
to their large lensing cross-sections (\citealt{TOG}). Compared with
their spiral and irregular counterparts, massive ellipticals are
known to host more GCs (e.g. \citealt{FMM1982, Harris1991,
Harris1993, West1993}). In this work, we consider the lensing effect
of GCs by imposing the observed Milky-Way GC population. If the lens
galaxies are much more massive than the Aquarius haloes, this
assumption may underestimate the appropriate GC abundance. However,
even with an increased number of GCs typical of massive ellipticals,
we would not expect major changes to our conclusions.

Thirdly, the semi-analytic galaxies that we have included in the
lensing simulation are all hosted by dark matter subhaloes above the
resolution limit of the Aquarius simulations. However, semi-analytic
models also follow the attached stellar components even after their
host dark matter (sub)haloes are tidally stripped below 20 particles
($\sim 10^5 h^{-1} M_{\odot}$). The baryonic masses of these
`sub-resolution' stellar components are in the same range as those
of satellite galaxies, and each is associated with a bound dark
structure no more massive than $10^5 h^{-1} M_{\odot}$. In {\sc
galform} their survival depends on an estimate of their merger
timescales, which is subject to the treatment of dynamical friction.
The radial distribution of the centres (most-bound particles) of
these objects is more concentrated than that of their counterparts
with resolved dark haloes. Nevertheless, including these
sub-resolution `orphan galaxies' only marginally increases the
cusp-violation probability under the assumption of singular
isothermal spheres for their density profiles.

Finally, when including semi-analytic galaxies to account for the
lensing effect of baryonic substructures, we have assumed that the
overall galactic potential and the dynamics of subhaloes would not
be significantly modified if baryons were added to our dark
matter-only simulations. However, a concentration of baryons at the
centre of a subhalo may increase its resilience to tidal stripping.
The abundance of subhaloes in the inner region of the main halo may
increase as a result of this effect. Our model does not account for
this `baryon-enhanced' survival of a small number of massive
objects, although we do not consider this omission to be
significant. There are likely to be few such objects, as a subhalo
must typically lose more than $\sim90\%$ of its mass before the
structure of its central (stellar) core influences further stripping
(see e.g. \citealt{Jorge08tidal}). This naturally limits the number
of `marginally destroyed' cases that would change in a fully
self-consistent model. Another effect of including baryons would be
to increase the concentration of the main halo in our simulation,
resulting in a stronger tidal field. However, studies of
disc-subhalo interactions using the Aquarius haloes (Lowing et al.
in prep) support our assumption that the subhalo population (in the
regime relevant to lensing) is only marginally affected by the
presence of a realistic stellar disc. Disc shocking (e.g.
\citealt{Kazantzidis2009,
  OnghiaSpringle2009DiskShocking}) may also reduce the inner
substructure abundance, which would further aggravate the problem of
explaining the observed lensing anomalies. \\

In \citet{Dandan09AquI}, the dark matter-only Aquarius simulations
were used to study the relation between dark matter substructure
abundance and lensing flux-ratio anomalies, in particular the
cusp-caustic violations observed for multiply imaged quasars. We
found that the dark substructures intrinsic to a typical
galaxy-scale lensing halo were not sufficient to explain the
observed frequency of cusp-caustic violations. In this work, we have
considered whether this expectation changes when satellite galaxies
in subhaloes are taken into account, or when the lensing halo is
assumed to contain a MW-like globular cluster population that is not
associated with surviving dark matter substructures. We have also
considered streams of dark matter identified in the Aquarius
simulations, in order to estimate the lensing effect of
irregularities in the halo itself. We conclude that the abundance of
intrinsic substructures, dark or bright, bound or diffuse, cannot
fully account for the observed cusp-violation frequency. Taken at
face value, this lack of substructure suggests a serious problem for
the CDM model. Warm dark matter models, which could reduce the
satellite abundance and may help to bring the dwarf galaxy LF into
agreement with observations without invoking strong feedback or
photoionization effects, (e.g. \citealt{SavalaWhite2010}) would only
make this problem worse. However, it is possible that the observed
frequency of flux anomalies are strongly biased by the small number
statistics of CLASS.

Previous studies have shown that intergalactic haloes ($M \lesssim
10^{10} M_{\odot}$) projected along the line of sight can cause
surface density fluctuations at the level of 1-10 per cent,
  making them a probable source of lensing flux anomalies
  \citep{Chen2003, Wambsganss2005, Metcalf2005a,
    Metcalf2005b,Miranda2007}. In future work, we will examine the
  contribution of this large-scale structure to observations of the
  cusp-caustic violation rate, using high-resolution cosmological
  volume simulations, which self-consistently model the halo mass
  function and clustering along the line of sight
  \citep[e.g.][]{PunchweinHilbert2009}.

\section*{Acknowledgments}
We thank Ian Browne, Neal Jackson and Simon White for useful
discussions and the referee for their comments. DDX has been
supported by a Dorothy Hodgkin fellowship for her postgraduate
studies. LG acknowledges support from a STFC advanced fellowship,
one-hundred-talents program of the Chinese Academy of Sciences (CAS)
and the National Basic Research Program of China (973 program under
grant No. 2009CB24901). CSF acknowledges a Royal Society Wolfson
Research Merit award. APC is supported by an STFC postgraduate
studentship. The simulations for the Aquarius Project were carried
out at the Leibniz Computing Centre, Garching, Germany; at the
Computing Centre of the Max Planck Society in Garching; at the
Institute for Computational Cosmology in Durham; and on the `STELLA'
supercomputer of the LOFAR experiment at the University of
Groningen.

\bibliographystyle{mn2e}
\bibliography{ms_xudd}
\label{lastpage}

\end{document}